# Application of CTS (Computer to Screen) Machine in Printing Industries for Process Improvement & Material Optimization

**Tarequl Islam** [1, *]

[1, *] Shahjalal University of Science and Technology, Kumargaon, Sylhet-3114, Bangladesh

Email: [1, *] tareq.ipe.sust@gmail.com

**Abstract:** The printing and labeling industries are struggling to meet the need for more complex and dynamic design requirements coming from the customers. It is now crucial to implement technological advancements to manage workflow, productivity, process optimization, and continual improvement. There has never been a time when the imagery and embellishments of apparel has been more commercially viable as it is now. Images and text are fused directly to fabric by heat transfer printing and labeling. For screen development which is required for heat transfer label mass production, many industries are still using the conventional method of screen development process. A CTS (computer-to-screen) innovates the printing and labeling industries by enhancing workflow, lowering consumable consumptions and chemical usage, speeding up setup, guaranteeing flawless design, and raising the print quality of the producing screens. The study's objective is to assess how CTS machines are used and how they affect existing heat transfer screen development processes in one of Bangladesh's leading printing and labeling companies. The study's primary goal is to highlight and analyze how the use of CTS machines reduces material and operational costs by optimizing the process. Costs for CapEx and OpEx are computed and compared for using CTS technology before and after adoption. Savings data such as material, consumable, and operating cost savings versus depreciation and machine payback period analysis were taken into consideration. It is clear from this study that CTS machines in the printing and labeling industries can guarantee profitability on top of Capital Expenditures.

## 1. Introduction

Competitiveness has been the key factor for the survival of the companies. The economic crisis that marked the beginning of this millennium forced the total readjustment of processes and operations which, in some cases, gave origin to deep changes in the organizations (1). The transformation trend towards digital technology to achieve sustainability targets and meet legal regulations has been visible in many industries. The printing sector has already been increasingly boosting sustainability performance through digitalization to automate workflows of processes (2). Also, the price of raw materials is increasing and manufacturers face cascading challenges through the supply chains (3). Thus the study will assess how adoption of CTS technology in a printing industry make the process optimized & remove a significant amount of materials consumption focusing on the Heat Transfer Screen development process.

Heat Transfer Printing & Labeling is one of the most important product lines in digital printing industries based on business contribution because it is very widely used in the apparel industry. Heat transfer labels convey the same information as their fabric equivalents – branding, washing information, size details etc. (4). A heat transfer is a method of taking a printed image and fixing the image directly to the garment in order to copy the image onto the garment itself. The design image is printed onto special transfer paper or synthetic film. This substrate has a special coating known as a release layer. This aids the transfer of the image onto the garment (5). When printed paper or synthetic film is pressed with a certain temperature & pressure, it provides tag-free strong brand story & information. Heat Transfer technologies follow a reference printed screen to be printed for large scale production. Each reference screen can be used for a certain amount of production run. To develop the reference screen as a prepress tool it is required to use artwork, design plate printing machine, silk screen, certain film, chemicals, exposure machine etc. & this is the conventional process of screen development. Whereas A CTS, or Computer-To-Screen printer, brings innovation to the screen printing industry. CTS machines do not use film; therefore, the cost of screen printing film is eliminated. Go straight from the computer to the emulsion-coated screen. By removing the need for film, image quality improves, and exposure times are dramatically reduced (6). T. Schweizer studied that using CTS technology in printing plate production helps to eliminate costly film material and expensive and environmentally harmful chemicals needed for developing the film (7).

The technology has been adopted by one of the major printing industries in Bangladesh. After successful launch of the project it is evident that CTS machine can improve screen development process and reduce screen development lead time ensuring quality & consistent design. The aim of the study is to understand how application of CTS machines have made a difference in the existing screen development process and make the process efficient ensuring optimized process & lowering material consumption.

## 2. Methodology

Adoption of this technology was one of the automation projects of Xyz printing & packaging ltd which is a market leader in the printing & labeling industry. After sufficient research and feasibility study, they decided to invest in a CTS machine in October, 2021 to replace the conventional process which was about to become obsolete. This was one of their global automation budgets. Background data were collected by collaborating directly with the project team. The existing system studied thoroughly & the impact of CTS technology also analyzed to make the project economically viable. Firstly, based on incoming customer demand the number of design plates required per day to produce a certain amount of reference screen to feed the machine was studied. Secondly, historical consumption of chemical & materials to fulfill the daily demand of reference screens were analyzed on a specific project timeline which were about to be consumed zero after CTS implementation. Required chemicals & materials cost had been analyzed & also impact on labor cost calculated as a savings amount. Machine purchasing cost, installation & training cost, freight cost & site preparation cost had been considered into investment amounts & also the electricity & depreciation cost after purchasing the machine had also been calculated. Based on the cost & savings analysis, financial data had been prepared about the IRR, NPV payback periods.

________________________________________________________________________________
Tarequl Islam[1, *]

After a financial feasibility study, it was found that the project had a small payback period with a significant IRR while ensuring a significant amount of savings.

**2.1 Current process of screen development**

For large scale production of heat transfer screens, every machine follows a prepressed & printed reference screen. Based on customer requirements the design has been created using any design tool such as Adobe Photoshop, Illustrator etc. Design file is loaded to the KATANA Imagesetter machine which requires time to develop & print out the positive plate. The plate is then mounted with a chemical coated mash or reference screen attached with red laser film (23'') so that the printing of a positive plate is exposed to the mash. Developer replenisher & fixer is used to attach the laser film with the screen & positive plate. Exposure machine is used to develop the mash through exposing, hardening & rinsing out by water. The heat transfer printer can print on areas that are not hardened so that ink can pass through and adhere to a surface. The reference or "mash" screen, which will be used to print screens at bulk level, is finally created.

They were making an average 108 positives plates for 180 screens per day and the positives making technology is in an obsolete stage and its maintenance cost is very high. Recently they had a breakdown with the KATANA Imagesetter machine and the fixing cost is about 7.5 K USD and machine parts are rare to find. For this they needed to outsource the positive and that costs about 250 USD daily. With the introduction of CTS Machine, they are being able to eliminate the positive development process. The investment will return within 3 years ensuring more than 70K USD saving per year.

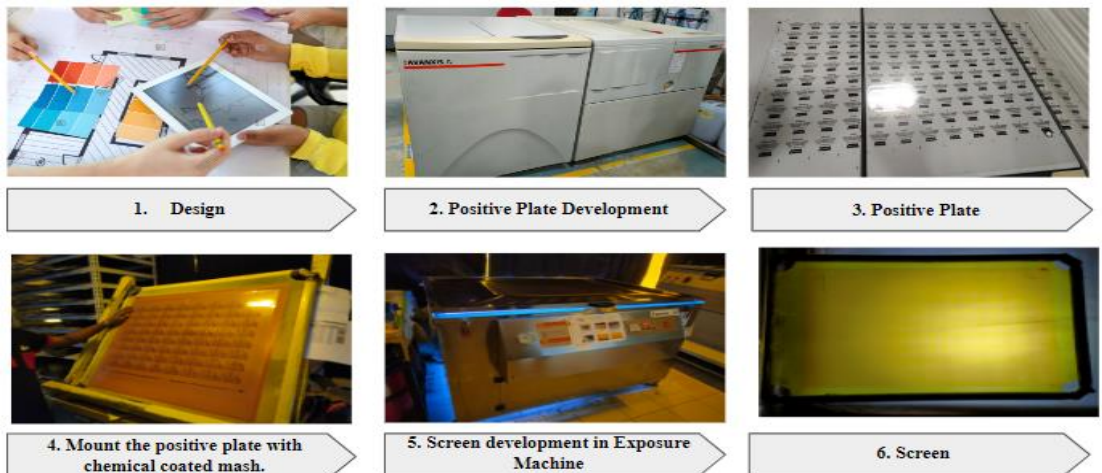

Fig1: Current screen development process

**2.2 Screen Development on CTS Machine**

CTS is a helpful device for imaging and exposing screens simultaneously needed to feed the printing machine for a specific impression sheet. The CTS machine does not require the chemical, laser film, positive plate development & exposure machine. The design file is uploaded to the CTS machine using RIP software. During the first pass over the screen frame, CTS uses specially formulated water-based UV blocking ink, CTS inkjet, and industrial pinheads. The emulsion-producing screen that can be transferred directly to washout is exposed on the return pass by the built-in high output scanning UV led light sources (8). The artwork is printed onto the screen by industrial print heads & simultaneously after printing the screen is already coated by UV led light. At the same time, the printed & exposed screen is hardened & rinsed out with water. Areas that the heat transfer screen printer wants to print are not hardened, so ink can pass through it which will be applied to the fabric surface through heat & pressure. The use of the Red Laser Film 23" roll, Developer Replenisher & Fixer is eliminated because of the adoption of CTS.

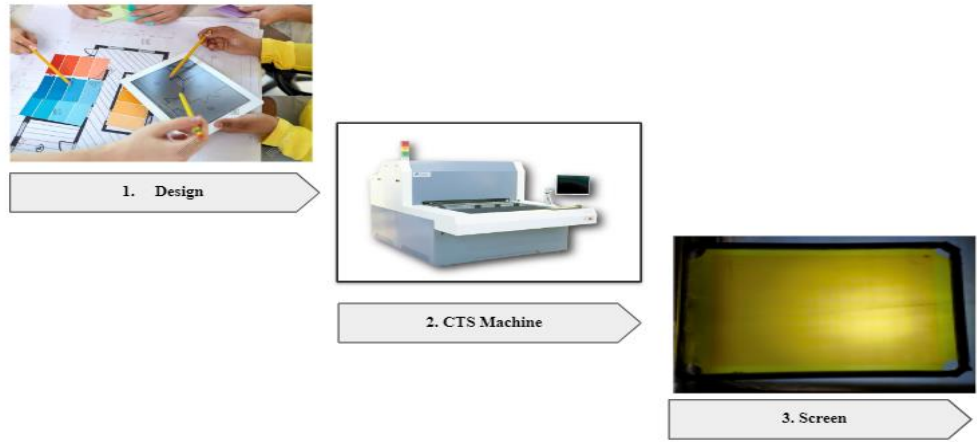

Fig2: Screen development on CTS machine

## 3. Historical consumption data

Below tables shows how chemical & laser film required for reference screen development per month before the technology adoption where unit costs for laser film is $252/roll & $4/Ltr for both replenisher & fixer-

Table1: Historical usage of chemical & material

| Month'21 | Film 23" (Roll) | Developer Replenisher (Ltr) | Fixer (Ltr) |
|---|---|---|---|
| January | 16 | 60 | 60 |
| February | 12 | 60 | 40 |
| March | 24 | 80 | 40 |
| April | 23 | 70 | 45 |
| May | 12 | 60 | 40 |
| June | 24 | 80 | 60 |
| July | 12 | 60 | 5 |
| August | 30 | 90 | 55 |
| September | 36 | 100 | 60 |
| Total | 189 | 660 | 405 |
| Total Price | $47,690 | $2,620 | $1,608 |
| **Annual Usage** | **$63,587** | **$3,494** | **$2,144** |

___________________________________________________________________________

Tarequl Islam[1,*]

## 4. Project investment summary

For successful CTS launching, the company had to invest a significant amount including direct & indirect costs. The investment cost includes the machine purchasing cost, installation & training cost, air freight cost & site preparation cost. Other two costs were analyzed & included in the total investment cost & these are depreciation cost of five fiscal years & day to day operating electricity cost. Below tables includes the cost breakdown of these particular projects:

Table2, 3,4: Machine investment cost, depreciation cost & electricity cost

(2)

| Project Investments of CTS Machine | |
|---|---|
| Machine Cost($) | $130,000 |
| Installation & Training Cost($) | $30,000 |
| Freight Cost($) | $5,000 |
| Site Preparation | $25,000 |
| **Total Cost($)** | **$190,000** |

(3)

| Depreciation Cost | |
|---|---|
| Number Of Machine Unit | 1 |
| Unit Price of CTS | $190,000 |
| Total Cost | $190,000 |
| Depreciation Year | 5 |
| **Total Depreciation Cost** | **$38,000** |

(4)

| Electricity Cost | |
|---|---|
| Total unit requirement (KW) | 4 |
| Average Run Hour | 20 |
| Efficiency | 70% |
| Electricity rate | $0.11 |
| Per Day Cost | $6 |
| **Total Electricity Cost** | **$1,785** |

## 5. Savings calculator

Below table includes the key deliverables & savings highlights from the project:

Table5: Key deliverables after adopting CTS machine

| Items | Current Cost (KUSD/Yr) | Future Cost (KUSD/Yr) | Saving (KUSD/Yr) | Deliverables | Highlights |
|---|---|---|---|---|---|
| Material | 63.5 | 0 | 63.5 | $ 63.5 K/Year | Eliminating laser film |
| Chemical | 5.6 | 0 | 5.6 | $ 5.6 K/ Year | Eliminating developer replenisher & fixer. |
| Head count | 1.6 | 0 | 1.6 | $ 1.6 K/ Year | Eliminating 0.5 HC from positive screen development. |

Breakdown of material, chemical & headcount savings data plotted in below tables:

_______________________________________________________________________________

Tarequl Islam[1,*]

Table6: Materials savings value (Yearly)

| Material Savings | |
|---|---|
| Red Laser Film 23" roll (Monthly) | $21 |
| Cost/ Roll | $252 |
| Monthly Savings | $5,299 |
| Total Annual SAVE | $63,587 |

Table7: Head count savings value (Yearly)

| HC Savings | |
|---|---|
| Annual Average CTC | $3,200 |
| Present HC | 1 |
| Future HC | 0.5 |
| Savings | $1,600 |
| Total Annual Savings | $1,600 |

Table8: Chemicals savings value (Yearly)

| Chemical Savings | | | | |
|---|---|---|---|---|
| Avg Monthly Developer Replenisher (Ltr) | 73 | Avg monthly use of Fixer (Ltr) | 45 |
| Cost / Ltr | $4 | Cost / Ltr | $4 |
| Monthly Savings | $291 | Monthly Savings | $179 |
| Yearly savings | $3,494 | Yearly savings | $2,144 |
| Total Annual Savings in Chemical ($) | | | | $5,637 |

## 5.1 Financial analysis summary

When the business case was raised, the finance team analyzed all the background data & found that the project has an internal rate of return of around 30% considering the best, worst & most likely scenario. The project had a payback period of around three years which indicates that adoption of this project can ensure profitability.

Table9: Financial analysis summary

| Particulars | Most Likely Scenario | Worst Case Scenario | Best Case Scenario |
|---|---|---|---|
| Project IRR | 30.3% | 30.2% | 30.4% |
| Net Present Value | $82,654 | $81,977 | $83,129 |
| Project Payback (yrs) | 3.06 | 3.06 | 3.05 |

## 6. Conclusion

Since October 2021, the company has significantly decreased the time and work needed to make the screens by adopting CTS technology. CTS screens are found superior to conventional screen development processes and they are offering more details and transition to meet customer preferences. With fast screen exposures, simple designs, consistent quality, and high speed, CTS has already reduced process and enhanced productivity in the Heat Transfer Label (HTL) department of the company. The space and maintenance needed for all the screens used every day has already decreased. Using CTS, they quickly print the reference screens and support the bulk production based on the needs of the production. Because customer preferences vary frequently and there are many variable types of information on the screens incoming these days, it is clear that adopting CTS made the process more efficient than the traditional screen development approach while ensuring profitability. It is recommended that any screen printing industries can adopt it as a part of their sustainable business operation based on their experience.

## Acknowledgement

I am grateful to the whole team of Xyz printing & packaging ltd who were directly or indirectly involved in the launching of this project & supported me with all the relevant information & guidance.

## References

1. Moreira, A., Silva, F., Correia, A., Pereira, T., Ferreira, L., & de Almeida, F. (2018). Cost reduction and quality improvements in the printing industry. Procedia Manufacturing, 17, 623-630. https://doi.org/10.1016/j.promfg.2018.10.107
2. Gladysz, B., Krystosiak, K., Ejsmont, K., Kluczek, A., & Buczacki, A. (2021). Sustainable Printing 4.0—Insights from a Polish Survey. Sustainability, 13(19), 10916. https://doi.org/10.3390/su131910916
3. Bragagni, Maurizio & Xhaferraj, Lorenc. (2021). RAW MATERIAL PRICES INCREASE. https://doi.org/10.13140/RG.2.2.10320.43526.
4. How to Print Labels Using Heat Transfers. Blog.focuslabel.com. (2016). Retrieved 19 August 2022, from https://blog.focuslabel.com/print-labels-using-heat-transfers.
5. Lee, D. (2019). What Is A Heat Transfer Label/Tag?. Blog.focuslabel.com. Retrieved 11 September 2022, from https://blog.focuslabel.com/what-is-a-heat-transfer-label.
6. Computer-To-Screen Machines | CTS Imaging. Lawson Screen & Digital Products. (2022). Retrieved 9 August 2022, from https://www.lawsonsp.com/screen-printing-equipment/auxiliary-equipment/cts-film-printers.
7. Schweizer, T. (2010). Computer-to-screen system handles large-scale applications. Https://Www.Researchgate.Net/, 32, 33-34. Retrieved 12 September 2022.
8. M&R i-Image XE Direct to Screen Machine - Screen Print World. Screen Print World. (2022). Retrieved 10 September 2022, from https://screenprintworld.co.uk/product/mr-i-image-xe-direct-to-screen-machine.